\documentclass{ws-p10x7}
\newcommand{\pom}{ $I\hspace{-1.6mm}P$}
\newcommand{\pome}{ I\hspace{-1.6mm}P}
\newcommand{\be}{\begin{equation}}
\newcommand{\ee}{\end{equation}}
\newcommand{\bea}{\begin{eqnarray}}
\newcommand{\eea}{\end{eqnarray}}
\newcommand{\bean}{\begin{eqnarray*}}
\newcommand{\eean}{\end{eqnarray*}}

\newcommand{\gapproxeq}{\lower
.7ex\hbox{$\;\stackrel{\textstyle >}{\sim}\;$}}
\newcommand{\lapproxeq}{\lower
.7ex\hbox{$\;\stackrel{\textstyle <}{\sim}\;$}}

\begin{document}
%\begin{titlepage}
%\begin{tabbing}
%wwwwwwwwwwwwwwwright hand corner using tabbing so it looks neat and in \= \kill
% \> {hep-ph/00xxxxx}   \\
% \> {RAL-00-xxx}   \\
% \> {BHAM-HEP/00-xx}   \\
%\> {sep 4 2001}
%\end{tabbing}
%\baselineskip=18pt
%\vskip 0.7in
%\begin{center}
\title{ Light Hadron Spectroscopy: Theory and Experiment}
%\vspace*{0.9in}
\author{Frank E. Close}
\address{ Dept of Theoretical Physics\\
Univ of Oxford, OX1 3NP, England\\e-mail: F.E.Close@rl.ac.uk} 
%\vspace{.1in}
%\\
%\vspace{0.1in}
%\vspace*{0.1in}
%\end{center}
%\maketitle
%\begin{abstract}
\twocolumn[\maketitle\abstract{
Rapporteur talk at the Lepton-Photon Conference, Rome, July 2001:
 reviewing the evidence and strategies for understanding 
scalar mesons, glueballs and hybrids, the gluonic Pomeron and the interplay of heavy 
flavours and light hadron dynamics.
Dedicated to the memory of Nathan Isgur, long-time collaborator
and friend, whose original ideas in hadron spectroscopy formed the basis for
much of the talk.  }]
%\end{abstract}
%\end{titlepage}
%\setcounter{page}{2}

\section{The Spectroscopy of Gluonic Excitations}
What is the nature of confinement, in particular the role of gluonic degrees of
freedom in the strong interaction limit of QCD? After decades of searches,
a new generation of precision data has transformed this field in the last
five years. In this talk I shall propose 
a potential solution to the problem of how glue is manifested and develop an experimental strategy to test it. The strategy involves a variety of complementary experiments,
ranging from $\gamma \gamma$ to electromagnetic production at Jefferson Lab and
other moderate energy dedicated machines, such as DA$\Phi$NE, to high energy
innovative experiments at RHIC, Fermilab and possibly even the LHC,which should
enable it to be refuted, refined or even substantially confirmed.
The scope of my talk will be to discuss
what theory expects, what the experimental situation is and what the new 
challenges for experiment are in light of these.   

Our main intuition on the strong interaction limit of QCD 
derives from Lattice QCD which implies a linear potential 
for $q\bar{q}$ systems\cite{Ba97}. This is well established for heavy flavours, where data confirm it. However, we are given a gift 
of Nature in that as one comes from heavy to light flavours, the linear potential 
continues phenomenologically to underpin the data: the S-P-D gaps are similar for 
$b\bar{b}$, $c\bar{c}$, and even $u\bar{d}$ as can be verified by looking in the 
PDG tables\cite{12gev,isgur1}. Even though we have no fundamental
understanding of why this is, we can nonetheless accept the gift and be confident that 
we can assign light flavoured mesons of given $J^{PC}$ to the required``slot" in the 
spectrum\cite{Go99}. I shall not review here the current status of $q\bar{q}$
spectroscopy as I am not interested in stamp collecting; all I wish to note at this point 
is that the resulting pattern leads one to expect that the lightest $J^{PC}$ $^3P_0$ 
$q\bar{q}$ nonet should occur in the region above 1 GeV - of which more in a moment.

Not everything is independent of flavour. The $^3S_1 - ^1S_0$ mass gap grows as the 
flavours get lighter, a phenomenon that is understood, at least qualitatively, from
 the chromomagnetic effects of gluon exchange\cite{12gev,isgur1}. Apart from this hint, gluonic effects 
have been all but absent - until now.

As the Coulomb potential of electrostatics implies that the electric fields tend to 
fill all three dimensions of space, so the linear potential implies that the analogous 
chromoelectric fields are highly collimated. This gave rise to the models where 
confinement arises from flux tubes\cite{fluxtubes,Is85}. 
There are now hints from the lattice\cite{Ba00} that 
such flux tubes indeed form, at least between static massive quarks. It is then 
almost model independent\cite{Lu81} that excitations of the flux tubes, analogous to phonons,
 will occur. The resulting new families are known as``hybrids"\cite{Is85,Cl95}. I shall
discuss the evidence in section 1.1

Lattice QCD predictions for the mass of the lightest (scalar) glueball are now mature.
In the quenched ,approximation the mass is 
$\sim 1.6$GeV\cite{Ba93,Se95,fcteper,Mo97,Wein}. Flux tube models
imply that if there is a $q\bar{q}$ nonet nearby, with the same $J^{PC}$ as the glueball,
then $G-q\bar{q}$ mixing will dominate the decay\cite{Am95}.
This is found more generally\cite{An98} and recent studies on coarse-grained
lattices appear to confirm that there is indeed significant mixing between 
$G$ and $q\bar{q}$
together with associated mass shifts, at least for the scalar sector\cite{McN00}.

Precision data on scalar meson production and decay are consistent with this and the
challenge now centres on clarifying the details and extent of such mixing. As
there have been several developments here recently, I shall devote much of my
talk to the question of scalar mesons, following the discussion of hybrids.

\subsection{Hybrid Mesons}

 Qualitatively the 
energy for their excitation will be of order $\frac{\pi}{R}$ where $R$ is the length 
scale between the quark coloured sources. This leads inexorably to the typical mass scale for light flavoured
hybrids at or just below 2GeV, with hybrid charmonium at $\sim$ 4 GeV.
Hybrid $D_g$ and $B_g$ heavy flavours will also exist but are beyond the scope of
this talk.

Hybrid states allow the existence of $J^{PC}$ correlations, such as 
$0^{+-},1^{-+},2^{+-}$ that are forbidden for conventional mesons. The flux-tube model
predicts that the lightest exotic hybrid, $J^{PC} = 1^{-+}$, lies just
below 2GeV\cite{Ba95,IKP85}; this is in accord with
the heuristic argument above and, perhaps more significantly, with 
lattice QCD\cite{Be97,Ju97,michael} for which  $1^{-+} s\bar{s}$ has 
mass $ 2.0 \pm 0.2$ GeV (and so the $n\bar{n}$ would be expected some 200 to
300 MeV lower). There should also be states that are gluonic excitations
 of the $\pi,\rho$, $0^{-+},1^{--}$\cite{bcd,Cl95}. During the last two years evidence for 
exotic $1^{-+}$ has emerged from more than one experiment and in various channels.

Two experiments\cite{e852,ves} have evidence for $\pi_1(1600) \equiv 1^{-+}$
in the $\rho^0\pi^-$ channel in the reaction $\pi^-N \to (\pi^+\pi^-\pi^-)N$.
At Serpukhov a 40 GeV/c $\pi^-$ beam shows this resonant $\pi_1(1600)$ also
in $\pi \eta'$ and $\pi b_1(1235)$, and at BNL the $\pi \eta'$ channel is also observed.

At BNL the channel $\pi f_1(1285)$ showed resonant $1^{-+}$ but nearer to 1800 MeV
in mass\cite{Cl95}. Resonant $1^{-+}$ also shows up in two experiments in the $\pi \eta$
channel\cite{e852b,cbexotic} but around 1400 MeV.

It is the proliferation of evidence that leads to the consensus that 
a resonant exotic state has been found. One possibility is that
the hybrid state is, in line with the lattice, $\sim 1.6 - 1.8$GeV
in mass and the $\pi f_1(1285)$ channel, which
opens up in S-wave at 1420 MeV with $1^{-+}$ quantum numbers, 
is playing some role in disturbing the phase shifts in the various channels.
These are details to be settled; the consensus is that an exotic $1^{-+}$
resonance is being manifested in different decay modes.

The question now is whether it is a true 
hybrid or more mundane, such as $qq\bar{q}\bar{q}$ or a molecular state of two mesons. 
To answer this question requires evidence in other channels - that is how we have 
historically determined the constituent nature of the ``traditional" mesons (after all,
 you could ask of any meson,``how can I tell if it is $q\bar{q}$ or a molecular combination of 
other mesons?"). The answers come when it is seen to be produced in a variety of processes,
 with common s-channel properties, and when the pattern of decays reveals its flavour 
content, in particular that it is not dominant in a single channel, as would be 
expected for a molecule. I shall return to this question later.

There are also tantalizing hints of  $\pi_g, 0^{-+}$ at 1.8 GeV and 
$\rho_g$, $1^{--}$ in the 1.45 - 1.7 GeV states. The former is seen 
prominently in unusual channels like 
$\pi f_0(980); \pi f_0(1500);KK^*_0; \pi \sigma$ and not in $\pi \rho; KK^*$;
 such patterns are as predicted for hybrids and contrast strongly with those 
expected of conventional mesons\cite{Cl95,nodes}. Adding further support
to the hypothesis that a hybrid $0^{-+}$ is being seen, is the evidence
\cite{vesb} for two isovector $0^{-+}$ states, $\pi(1600)$ and $\pi(1800)$
where the quark model predicts only one.

 Whatever the $\pi(1800)$ is, its existence should 
concern anyone who is interested in D-decays. The D and $\pi_g$ are degenerate 
and so the Cabibbo suppressed decays of the D may be affected\cite{fclipkin}.
 It is therefore 
important to determine the pattern of decays of the $\pi_g$ and to compare with
 the Cabibbo suppressed modes of the D; e.g. the prominent $\pi \sigma$
in $D^+ \to \pi^- \pi^+ \pi^+$\cite{e791b}. Similar remarks may apply to the $D_s$ 
and the (as yet unseen) $K_g$ strange partner.

The phenomenology of the vector mesons in the 1.45-1.7 GeV region seems to 
require hybrid content. One of the driving features is that a conventional 
$1^{--}$ has the $q\bar{q}$ with S=1, whereas for a hybrid it is the gluonic 
degrees of freedom that give the net J=1 while the $q\bar{q}$ are coupled to spin zero.
 This different internal structure gives rise to characteristic properties, which appear 
to be realized in the data\cite{donnachie,Cl95}. This should be a source of active future study
 for DA$\Phi$NE or VEPP. Conversely, the exotic $1^{-+}$ has $S_{q\bar{q}} = 1$ and so 
there is the possibility that photons can produce the $q\bar{q}$ conveniently in the
 spin triplet state; $\omega$ or $\pi$ exchange at low energies can be an entree into 
the hybrid sector. Thus photo and electroproduction at Jefferson Lab, in their 6 GeV 
programme and especially at a 12 GeV upgrade, will be exciting\cite{Is85,Cl95,12gev}. 

\subsection{Glueballs}

The third, and most well advertised, aspect of glue is the prediction that there 
exist glueballs (or in flux tube language, glue loops)\cite{fglglue}. The lattice has matured and
 converged on the following. The lightest glueball, in the quenched approximation, 
is predicted to be $0^{++}$ with a mass in the 1.4 - 1.7 GeV region; the next lightest
 being $2^{++}, 0^{-+}$ at around 2 GeV\cite{Ba97,Ba93,Se95,fcteper,Mo97}.
 There has been considerable progress in this 
area, especially as concerns the scalar glueball.

A problem is that the maturity of the $q\bar{q}$ spectrum, as already mentioned, 
tells us that we anticipate the $0^{++}, q\bar{q}$ nonet to occur in the 1.3 to 1.7 
GeV region. Any such states will have widths and so will mix with a scalar glueball 
in the same mass range. It turns out that such mixing will lead to three physical 
isoscalar states with rather characteristic flavour content\cite{fcteper,ClK01a}.
 Specifically; two will
 have the $n\bar{n}$ and $s\bar{s}$ in phase (``singlet tendency"), their mixings with
 the glueball having opposite relative phases; the third state will have the $n\bar{n}$ 
and $s\bar{s}$ out of phase (``octet tendency") with the glueball tending to decouple in 
the limit of infinite mixing. There are now clear sightings of prominent scalar 
resonances $f_0(1500)$ and $f_0(1710)$ and, probably also, $f_0(1370)$.
(Confirming the resonant status of the latter is one of the critical pieces
needed to clinch the proof - see ref.\cite{klempt} and later).  The 
production and decays of these states is in remarkable agreement with this flavour 
scenario\cite{ClK01a}.

Were this the whole story on the scalar sector there would be no doubt that the 
glueball has revealed itself. However, there are features of the scalars in the 
region from two pion threshold up to $O(1)$GeV that have clouded the issue. There
 has been considerable recent progress here that enable a consistent picture to be 
proposed. I will now set the scene for this and return later to the experimental 
challenges.

\section{Scalar Mesons: Above and Below 1 GeV}

To understand the message of the $J=0^{++}$ data one must distinguish 
the regions above and below 1 GeV, typified by the $K\bar{K}$ threshold.

The $q\bar{q}$ $^3P_0$ nonet is expected in the 1.3 - 1.7 GeV mass
region. There is empirically a nonet (at least) and we shall return to this 
later. If that was the whole story then the scalar spectroscopy would be 
understood. But there are also the $f_0(980)$ and $a_0(980)$ mesons,
and possibly a $\kappa$ and $\sigma$ below 1 GeV.
(While the $\sigma$ is claimed in recent data
 in charm decays\cite{e791b}, there are conflicting
conclusions about the $\kappa$\cite{kappadata}; a theoretical
critique is in ref.\cite{kappa}).
 Various attempts have been made to
describe them as a $q\bar{q}$ nonet shifted to low mass by some mechanism. 
While one cannot formally disprove such a picture, it ignores a fundamental
aspect of QCD that appears to be realized in the data. 

As pointed out by Jaffe\cite{Ja77} long ago, there is a strong QCD attraction among $qq$ 
and $\bar{q}\bar{q}$ in S-wave, $0^{++}$, whereby a low lying nonet of scalars may be 
expected. As far as the quantum numbers are concerned these states will be like 
two $0^{-+}$ $q\bar{q}$ mesons in S-wave. In the latter spirit, Isgur and Weinstein\cite{iswein} had 
noticed that they could motivate an attraction among such mesons, to the extent 
that the $f_0(980)$ and $a_0(980)$ could be interpreted as $K\bar{K}$ 
molecules. 

The relationship between these is being debated
\cite{tornqv,penn,achasov2,achasov1,speth}, and I shall contribute to it
here, but while the details may be argued about,
there is a rather compelling message of the data as follows.
 Below 1 GeV the phenomena 
point clearly towards an S-wave attraction among two quarks and two antiquarks
(either as $(qq)^{\bar{3}}(\bar{q}\bar{q})^3$, or $(q\bar{q})^1(q\bar{q})^1$ where superscripts
denote their colour state), while
above 1 GeV it is the P-wave $q\bar{q}$ that is manifested. There is a critical
distinction between them: the 
``ideal" flavour pattern of a $q\bar{q}$ nonet on the one hand,
and of a $qq\bar{q}\bar{q}$ or meson-meson,
 nonet on the other, are radically different; in effect they
 are flavoured inversions of one another. Thus whereas the
former has a single $s\bar{s}$ heaviest, with strange in the middle and and I=0; I=1
set lightest (``$\phi;K;\omega,\rho$-like"), the latter has the I=0; I=1
set heaviest ($K\bar{K};\pi\eta$ or $s\bar{s}(u\bar{u} \pm d\bar{d}$)) with strange in the
middle and an isolated I=0 lightest ($\pi\pi$ or $u\bar{u}d\bar{d}$)\cite{Ja77,iswein,Jaf00}.

The phenomenology of the $0^{++}$ sector appears to exhibit both of these patterns with
$\sim 1$GeV being the critical threshold. Below 1 GeV the inverted structure of the four quark
dynamics in S-wave is revealed with $f_0(980);a_0(980)$; $\kappa$ and $\sigma$ as the labels.
One can debate whether these are truly resonant or instead are 
the effects of attractive long-range $t-$channel dynamics
 between
the colour singlet $0^{-+}$ $K\bar{K}; K\pi; \pi\pi$, but the systematics
of the underlying dynamics seems clear. Above 1 GeV the $^3P_0$ $q\bar{q}$ nonet should
be apparent: there are candidates in $a_0(\sim 1400);f_0(1370);K(1430);f_0(1500)$ and $f_0(1710)$.
One immediately notes that if all these states are real there is an excess, precisely as
would be expected if the glueball predicted by the lattice is mixing in this region. 

A major question is whether the effects of the glueball are localised in this region above 1 GeV,
as discussed by ref\cite{FC00,ClK01a} or spread over a wide range, perhaps down to the
$\pi\pi$ threshold\cite{minkochs}. This is the phenomenology frontier. There are also two
particular experimental issues that need to be settled: (i) confirm the existence of $a_0(1400)$ and
determine its mass (ii) is the $f_0(1370)$ truly resonant or is it a $t-$channel exchange phenomenon 
associated with $\rho\rho$\cite{klempt}.

As concerns the region below 1GeV, the debate centres on whether the phenomena are
truly resonant or driven by attractive t-channel
exchanges, and if the former, whether they are 
molecules or $qq\bar{q}\bar{q}$. These are, in my opinion, secondary issues; 
each points to the strong
attraction of QCD in the scalar S-wave nonet channels. The difference between 
molecules and compact $qq\bar{q}\bar{q}$ will be revealed in the
tendency for the former to decay into a single dominant channel - the molecular constituents -
while the latter will feed a range of channels driven by the flavour spin clebsch gordans. This is a general means to decide whether an exotic
$1^{-+}$ is a hybrid or a molecule, for example, and for the light
scalars has its analogue in the production characteristics. 

The picture that is now emerging from both phenomenology\cite{ClK01b,ClK01c,kloe} 
and theory\cite{jaffe01} is that
both components are present. As concerns the theory\cite{jaffe01}, 
think for example of the two component picture as two channels. One, the quarkish channel 
($QQ$) is
somehow associated with the $(qq)^{\bar 3}(\bar{q}\bar{q})^{3}$ coupling of a two quark-two
antiquark system, and is where the attraction comes from. The other, the meson-meson channel
($MM$) could be completely passive (eg, no potential at all). There is some
off diagonal potential which flips that system from the  $QQ$  channel to $MM$. 
The way the object appears to experiment depends on the strength of the
attraction in the $QQ$ channel and the strength of the off-diagonal
potential. The nearness of the $f_0$ and $a_0$ to $K\bar{K}$ threshold suggests
that the $QQ$ component cannot be too dominant, but the fact that
there is an attraction at all means that the $QQ$ component cannot be 
negligible. So in this line of argument, $a_0$ and $f_0$ must be
four-quark states and $K\bar{K}$ molecules at the same time.

If one adopts this as a reference hypothesis, many data begin to
make sense. I shall discuss this in section 4; first I
shall concentrate on the glueball search, to decide if
a scalar glueball is manifested above 1GeV.

\section{Glueball production dynamics}

The folklore has been that to enhance glueball signals one should concentrate on 
production mechanisms where quarks are disfavoured: thus $\psi \to \gamma G$\cite{chan},
$p\bar{p} \to \pi + G$ in annihilation at rest\cite{chan,fcrpp}, and central production
in diffractive (gluonic pomeron) processes, $pp \to p G p$\cite{robson,fcrpp}. 
Contrasting this, 
$\gamma \gamma$ production should favour flavoured states such as $q\bar{q}$.
Thus observing a state in the first three, which is absent in the latter,
would be prima facie evidence.

Such ideas are simplistic. There has been progress in quantifying them
and in the associated phenomenology. The central production has matured
significantly in the last three years and inspires new experiments
at RHIC, Fermilab and possibly even the LHC. Thus I shall concentrate
on this, but first show how these complementary processes collectively are
now painting a clearer picture.

First on the theoretical front, each of these has threats and opportunities.
(i)In $\psi \to \gamma G$ the gluons are timelike and so it is reasonable
to suppose that glueball will be favoured over $q\bar{q}$ production.
Quantification of this has been discussed in ref.\cite{cfl} with some
tantalising implications: (a) the $f_0(1500)$ and $f_0(1710)$ are produced with
strengths consistent with them being $G-q\bar{q}$ mixtures, though there are
some inconsistencies between data sets that need to be settled 
experimentally; (b) the $\xi(2200)$ is produced with a strength consistent
with that for a glueball, with two provisos: - that it has spin 2 
(which is probably the case), and that it really exists (which is debatable, see section 5).

(ii) In $p\bar{p}$ the $q$ and $\bar{q}$ can rearrange themselves to produce
mesons without need for annihilation. So although a light glueball may
be produced, it will be in competition with conventional mesons and
any mixed state will be produced significantly by its
$q\bar{q}$ components.

(iii) In central production the gluons are spacelike and so must rescatter
in order to produce either a glueball or $q\bar{q}$. Thus here again one
expects competition. However, a kinematic filter has been discovered\cite{ck97}, which
appears able to suppress established $q\bar{q}$ states, when the $q\bar{q}$
are in P and higher waves.

Its essence was that the pattern of resonances produced in the central 
region of double tagged $pp \rightarrow pMp$ depends on the vector 
$ difference$ of the transverse momentum recoil of the final state
protons (even at fixed four momentum transfers). When this quantity
($dP_T \equiv |\vec{k_{T1}} - \vec{k_{T2}}|$) is large, ($\geq O(\Lambda
_{QCD})$), $q\bar{q}$ states are prominent whereas at small
$dP_T$ all well established $q\bar{q}$ are observed to be suppressed
while the surviving resonances include the enigmatic $f_0(1500),
f_0(1710)$ and $f_0(980)$.

The data are consistent with the hypothesis that as $dP_T \rightarrow 0$
all bound states with internal $L > 0$ (e.g. $^3P_{0,2}$ $q\bar{q}$)
are suppressed while S-waves survive (e.g. $0^{++}$ or $2^{++}$ glueball 
made of vector gluons and the $f_0(980)$ as any of glueball,
or S-wave $qq\bar{q}\bar{q}$ or $K \bar{K}$ state). Models are
needed to see if such a pattern is natural.
As the states that survive this cut appear to have
an affinity for S-wave, this may be evidence for $qq\bar{q}\bar{q}$ 
or $q\bar{q}q\bar{q}$ (as for example the $f_0(980)$) or for $gg$ content
(as perhaps in the case of $f_0(1500;1710)$ and $f_2(1930)$). It would be interesting
to study the production of known $q\bar{q}$ states in $e^+ e^- \to e^+ M e^-$
to see how they respond to this kinematic filter, and gain 
possible insights into its dynamics.

\begin{figure}
\epsfxsize120pt
\figurebox{120pt}{160pt}{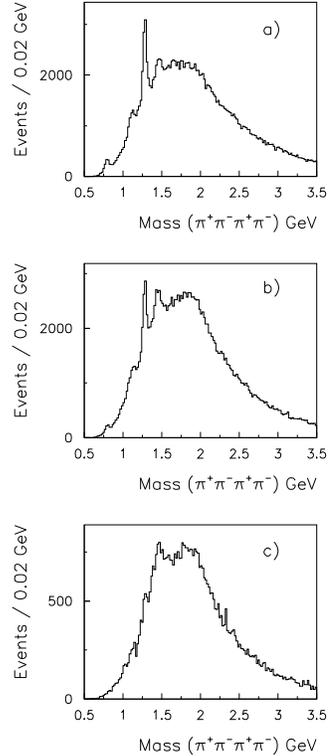}
%\centerline{\epsfig{file=dptt2.eps,width=3.0in,angle=0}}
\caption{ $pp \to pp + (2\pi^+2\pi^o)$.
As $dP_T$ decreases from (a) $>0.5 GeV$ to (b) $0.2<dP_T<0.5$ GeV and (c) $<0.2$ GeV 
the $^3P_1(q\bar{q})f_(1285)$ in (a) disappears while the $f_0(1500)$
and $f_2(1930)$ glueball candidates become prominent.\label{fig:dpt}}
\end{figure}

Following this discovery there has been an intensive experimental
programme in the last three years by the WA102 collaboration at CERN, which
has produced a large and detailed set of data on both the
$dP_T$
~\cite{ck97} and the azimuthal angle,
$\phi$,
dependence of meson 
production (where 
$\phi$ is the angle between the transverse momentum
vectors, $p_T$, of the two outgoing protons).
\par
The azimuthal dependences 
as a function of
$J^{PC}$ and the momentum transferred at the proton vertices, $t$,
are very striking. As seen in refs.~\cite{WAphi}, and
later in this talk, the $\phi$ distributions for mesons with
$J^{PC}$~=~$0^{-+}$ maximise around $90^{o}$, $1^{++}$ at $180^o$ and $2^{-+}$
at $0^o$. This is not
simply a J-dependent effect~
\cite{pi4papr} since
$0^{++}$ production peaks at $0^o$ for some states whereas others are more
evenly spread~\cite{WA0++}; $2^{++}$ established $q\bar{q}$ states peak at
$180^o$ whereas the $f_2(1950)$, whose mass may be consistent with the tensor
glueball predicted in lattice QCD, peaks at $0^o$~\cite{pi4papr} (see also
fig. 5 in the present paper).
These data are all explained if the Pomeron transforms as a non-conserved
vector current: specifically, having an intrinsic and important scalar component.
 The detailed calculations are described in~\cite{cs1,cs2}.
Production of the $0^{-+},1^{++},2^{-+}$ sequence
and the absence of a $0^{-+}$ glueball candidate in these
data are now understood; the $0^{++},2^{++}$ states, where
$q\bar{q}$ and glueballs are expected, are tantalizing. I shall
now survey these $J^{PC}$ states in turn.

\noindent $J^{PC}=0^{-+}$

Here I shall concentrate on the $\eta^\prime$ meson
whose production has been found to be consistent with double
pomeron exchange~\cite{WAphi}. The resulting
behaviour of the cross section may be summarised as follows:

\[
\frac{d\sigma}{dt_1 dt_2 d \phi }  
\sim t_1 t_2 {G^p_E}^{2} (t_1)
{G^{p}_{E}}^2 (t_2) \sin^2(\phi)\]
\[ \times F^2(t_1, t_2, M^2) 
\]
% \noindent where $F(t_1, t_2, M^2)$ is the \pom-\pom-$\eta^\prime$
\noindent where $\phi $ is the angle between the two $pp$ scattering
planes in the \pom-\pom\thinspace centre of mass.
  $pp$ elastic scattering data
and/or a Donnachie Landshoff type form factor~\cite{dl} can be used as
model of the proton-\pom\thinspace form factor ($G_E^p(t)$).
  $F(t_1, t_2, M^2)$ is the \pom-\pom-$\eta^\prime$
form factor, 
parametrised as $exp^{- b_T(t_1+t_2)}$ where the sole parameter $b_T$~=~0.5~$GeV^{-2}$
in order to describe the $t$
dependence.
Fig.(2) compare the final theoretical form for the $\phi$
distribution and the $t$ dependence with the data for
the $\eta^\prime$.

\begin{figure}
\epsfxsize200pt
\figurebox{200pt}{160pt}{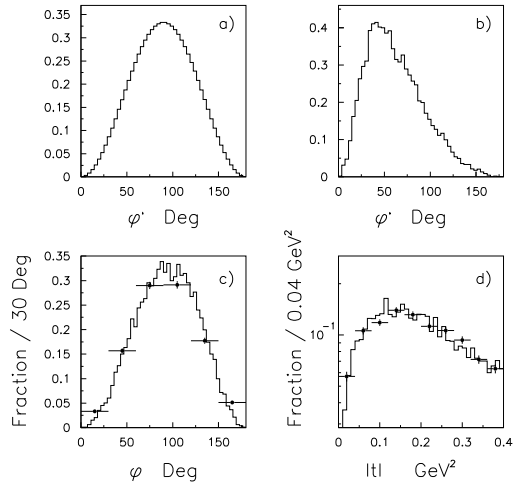}
\vspace{-.5cm}
\caption{ The predicted $\phi$ distributions for $J^{PC}$~=~$0^{-+}$ mesons
a) naive distribution and b) taking into account the experimental kinematics.
c) The $\phi$
and d) the $|t|$ distributions for the $\eta^\prime$
for the data (dots) and the model predictions from the Monte Carlo (histogram).\label{fig:2}}
\end{figure}

Parity requires the vector pomeron to be transversely polarized, which
gives rise to the $t_1t_2$ factors in the cross section. For
$0^{-+}$ states with $M >> 1 $GeV, as expected for the lattice glueball or radial
excitations of $q \bar{q}$, this dynamical $t_1t_2$ factor will suppress 
the region where kinematics would favour the production.  It would be interesting if 
glueball production dynamics involved a
singular $(t_1t_2)^{-1}$ that compensated for the transverse \pom factor, as
in this case the cross section would be enhanced. However, we have no reason
to expect such a fortunate accident, so observation of high mass
$0^{-+}$ states is expected to be favourable only at higher energies, such as at RHIC, Fermilab
or LHC.

\noindent $J^{PC} = 1^{++}$

Refs.~\cite{cs1,cs2,fc1+} predicted that axial mesons
are produced polarised, dominantly in helicity one, from one Pomeron that is
polarized transversely and one longitudinally. This is verified by
data
\cite{WA1+pol}.  Such spin dependence leads to a cross section structure

\[
\frac{d\sigma}{dt_1dt_2d\phi } 
\sim
[(\sqrt{t_1} - \sqrt{t_2})^2 + 4 \sqrt{t_1t_2}
\sin^2(\phi /2)]
\]
\[ \times F^2(t_1,t_2,M^2) 
\]

\noindent which implies a dominant $\sin^2(\phi/2)$ behaviour that tends to
isotropy when suitable cuts on $t_i$ are made. This is qualitatively
realized.

Fig.~(3) show the output of the model predictions from the
Galuga Monte Carlo superimposed
on the $\phi$ and $t$ distributions for the $f_1(1285)$ from
the WA102 experiment; the
(parameter-free)
prediction of
the variation of the $\phi$ distribution as a function of
$|t_1-t_2|$ is shown in
Fig.~(3c and d).
The agreement between the data and
theory is excellent.
Similar conclusions arise for the $f_1(1420)$.

\begin{figure}
\epsfxsize200pt
\figurebox{200pt}{160pt}{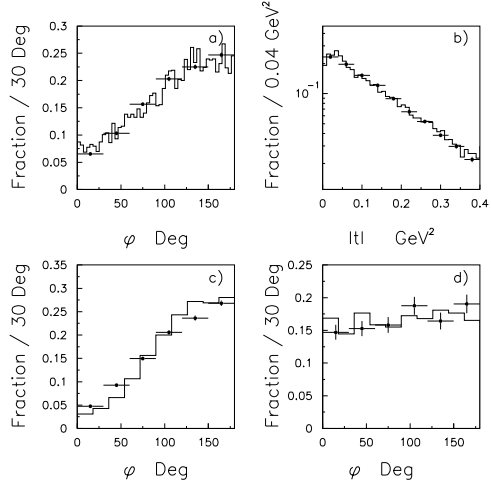}
\vspace{-.5cm}
\caption{ a) The $\phi$
and b) the $|t|$
distributions for the $f_1(1285)$
for the data (dots) and the Monte Carlo (histogram).
c) and d) the $\phi$ distributions for $|t_1 - t_2|$~$\le$~0.2 and
$|t_1 - t_2|$~$\ge$~0.4~$GeV^{2}$ respectively.\label{fig:3}}
\end{figure}

In passing, the resulting analysis of axial meson production\cite{Cl97} implies that
the $f_1(1285;1420)$ are members of a nonet with 

\begin{eqnarray}
f_1(1285) \sim |n\bar{n}\rangle - 0.5 |s\bar{s} \rangle \nonumber \\
f_1(1420) \sim |s\bar{s} \rangle + 0.5|n\bar{n}\rangle \nonumber 
\end{eqnarray}
\noindent ($n\bar{n} \equiv (u\bar{u} + d\bar{d})/\sqrt{2}$).
DELPHI\cite{delphi} have measured the production rate $<n>$ of these states
per hadronic  $Z$ decay and find $<n(1285)> = 0.13 \pm 0.03; <n(1420)> = 0.05 \pm 0.01$.
These are typical of the $<n>$ for states with $n\bar{n}$ content, supporting the
conclusion\cite{Cl97} that these mesons are partners in a nonet, each with non-negligible
$n\bar{n}$ content. L3 also report at this conference\cite{L3} the $\gamma \gamma$
couplings to these states, in accord with these conclusions.

\noindent $J^{PC}=2^{-+}$

The $J^{PC}$~=~$2^{-+}$ states, the $\eta_2(1645)$ and $\eta_2(1870)$,
are predicted to be produced
polarised. Helicity 2 is suppressed by Bose symmetry~\cite{cs1} and
has been found to be negligible experimentally~\cite{WA2-+}.
The structure of the cross section
is then predicted to be dominantly in helicity-one.
The helicity zero distribution is  as for the $0^{-+}$ case,
\[
\frac{d\sigma}{dt_1dt_2d\phi } \sim t_1t_2 \sin^2(\phi)
\]

\noindent and hence suppressed for $M \geq 1 GeV$ as was the case for the $0^{-+}$.
The dominant helicity-one distribution
is the same as for the $1^{++}$ case except for the important and
significant change from $\sin^2(\phi/2)$ to $\cos^2(\phi/2)$.

The results of the WA102 collaboration for the $\eta_2(1645)$~\cite{WA2-+}
are shown
in fig.~(4a and b).
The distribution peaks as $\phi \to 0$, in
marked contrast to the suppression in the $1^{++}$ case (fig.~3a);
(the $\eta_2(1870)$ results are qualitatively similar).

\begin{figure}
\epsfxsize200pt
\figurebox{200pt}{160pt}{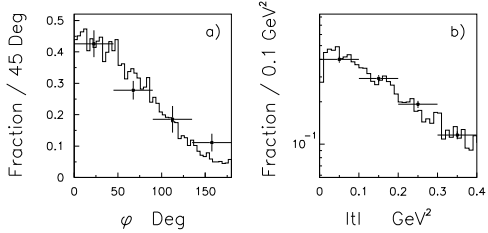}
\vspace{-3.5cm}
\caption{ a) The $\phi$
and b) the $|t|$
distributions for the $\eta_2(1645)$
for the data (dots) and the Monte Carlo (histogram).\label{fig:4}}
\end{figure}

\noindent Bearing in mind that there are no free parameters, the
agreement is remarkable. Indeed, the successful
description of the $0^{-+}$, $1^{++}$ and $2^{-+}$ sectors, both
qualitatively and in detail, sets the scene for our
analysis of the $0^{++}$ and $2^{++}$ sectors where glueballs are
predicted to be present together with established $q\bar{q}$ states.

\noindent $J^{PC}=0^{++}$ and $2^{++}$

Here we
 find that the production topologies do depend on
the internal dynamics of the produced meson and as such may
enable a distinction between $q\bar{q}$ and exotic, glueball, states.

In contrast to the $0^{-+}$ case, where parity forbade the LL amplitude,
in the $0^{++}$ case both $TT$ and $LL$ can occur. Hence there are two
independent form factors~\cite{cfl} $A_{TT}(t_1,t_2,M^2)$ and
$A_{LL}(t_1,t_2,M^2)$. For
$0^{++}$ and the helicity zero amplitude of $2^{++}$ (which
experimentally is found to dominate~\cite{WAhel0}) the angular dependence
of scalar meson production has the form~\cite{cs2}

\[
\frac{d\sigma}{d \phi}
\sim  [1 + \frac{\sqrt{t_1t_2}}{\mu^2}a_T/a_L 
\cos(\phi)]^2 
\]

\noindent which successfully predicts that there should be significant changes
in the $\phi$ distributions as $t$ varies~\cite{WAphi}.

The overall $\phi$
dependences for the $f_0(1370)$, $f_0(1500)$, $f_2(1270)$ and
$f_2(1950)$ can be described by varying the quantity $\mu^2a_L/a_T$.
Results are shown in fig.~5.
It is clear that these $\phi$
dependences discriminate two classes of meson in the $0^{++}$ sector and
also in the $2^{++}$.
The $f_0(1370)$ can be described using $\mu^2a_L/a_T$~=~-0.5~$GeV^2$,
for the $f_0(1500)$ it is +0.7~$GeV^2$,
for the $f_2(1270)$ it is -0.4~$GeV^2$ and
for the $f_2(1950)$ it is +0.7~$GeV^2$.

\begin{figure}
\epsfxsize200pt
\figurebox{200pt}{160pt}{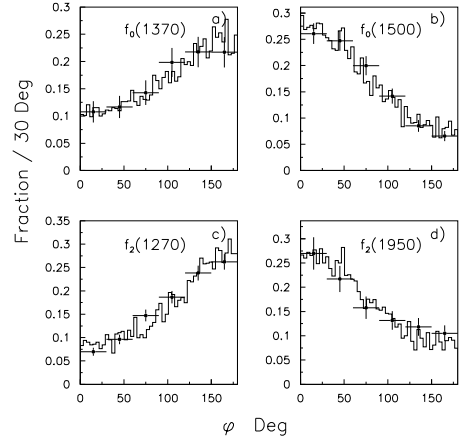}
\vspace{-.5cm}
\caption{The $\phi$
distributions for the a) $f_0(1370)$,
b) $f_0(1500)$, c) $f_2(1270)$ and d) $f_2(1950)$
for the data (dots) and the Monte Carlo (histogram).\label{fig:5}}
\end{figure}

It is interesting to note that these $\phi$ distributions can be fitted
with just one parameter and it is primarily the sign of this quantity that
drives the $\phi$ dependences.
Understanding the dynamical origin of this sign is now a central challenge
in the quest to
distinguish $q \overline q$ states from
glueball or other exotic states.

\section{Heavy Flavours and Light Scalars}

 The working hypothesis is that the $q\bar{q}$ nonet, mixed with a glueball, is realised
above 1 GeV and we now need to determine the
flavour content of these states. In addition we need
to confirm the picture of the $f_0(980)$ and
light scalars below 1 GeV.

Novel probes are provided by the decays of heavy flavoured states. They
can provide important information about the enigmatic $0^{-+}$ and $0^{++}$
light flavoured states in particular. The nature of the $\eta-\eta'$ system
remains an important open question with an impact on heavy flavour
decays. The $D$ and $B$ meson decays into final states containing $\eta$
or $\eta'$ continue to challenge theoretical predictions.
Further experiments can test the nature of these $0^{-+}$ states, e.g.
$D^+$ versus $D_s \to \eta(\eta') e^+ \nu$ and $B^o$ versus
$B_s \to J/\psi \eta(\eta')$\cite{lipkin1}.

$\psi \to \gamma \pi\pi$ compared with $D_s \to \pi \pi \pi$, and
$\psi \to \gamma K\bar{K}$ compared with $D_s \to \pi K\bar{K}$
provide complementary entrees into the light flavoured $0^{++}$ mesons.  Comparison with
$D_d \to \pi K^*_0(1430)$ then enables
us to ``weigh" the flavour content of the nonets. In
$\psi \to \gamma K\bar{K}$ Dunwoodie\cite{dunwoodie} finds the $f_0(1710)$ as clear
scalar, and this state could be sought in $D_s \to \pi K\bar{K}$
with enough statistics (in E687\cite{e687} 
the $K^*\bar{K}$ band contaminates the 1710 region of the Dalitz plot). 
The momentum transfer in $D_s \to \pi f_0(1710)$ is small enough that a non-relativistic
calculation of the transition amplitude $\langle c\bar{s} (^1S_0)|  \pi (\vec{k}) |
s\bar{s} (^3P_0) \rangle$ may be reliable. In particular this could distinguish between 
the $f_0(1710)$ as a radially excited state, for which there is significant suppression,
and a pure $ s\bar{s} (1^3P_0)$ for which the rate would maximise. Quantifying this rate
could be a challenge for high statistics data e.g. with
FOCUS. 
 The major signal in the E687 data is
the $\phi$; the $f_0(980)$ is just below threshold and it is not discussed whether
any of the signal at threshold is due to this state. However, in the E791 data\cite{e791} on
$D_s \to \pi \pi \pi$ the $f_0(980)$ is very prominent, together with the $f_0(1370)$
and a possible (though unclaimed) hint of a shoulder that could 
signal the $f_0(1500)$. Dunwoodie's analysis of $\psi \to \gamma \pi\pi$
shows structure around 1400 MeV and with better statistics from BES and
Cornell this should be verified and attempts to resolve it into $f_0(1370)$
and $f_0(1500)$ made. The strength of $f_0(1710)$ in these data should also 
be determined. 

One clear message from the E791 data is that $f_0(980)$ has strong affinity for
$s\bar{s}$ in its production; it decays into $\pi\pi$ as the $K\bar{K}$ channel
is closed. This brings us naturally to new information, presented to this
conference, which may at last help to solve the conundrum of the nature of
the $f_0(980)$.

We argued earlier that the $f_0(980)$ and $a_0(980)$ have strong affinity
for a four-quark composition, the question at issue being whether they are
compact $qq\bar{qq}$ or meson molecules. The emerging picture from 
phenomenology\cite{ClK01b,ClK01c,kloe} and theory\cite{jaffe01} is that
both components are present. 

There is a large amount of data on the production of the $f_0(980)$ which
require in some cases a strong affinity for $s\bar{s}$ (e.g. the $D_s$ decays
already ), or for $n\bar{n}$ (the production in hadronic Z decays has
all the characteristics associated with well established $n\bar{n}$ states\cite{delphi,lafferty}) and also data that require both components to be
present ($\psi \to  \omega f_0$ versus $\psi \to \phi f_0$). There are new data
that touch on the relationship between these states.

First I summarise arguments that the $f_0$ and $a_0$ are mixed. Then
I shall review ideas on $\phi \to \gamma f_0/\gamma a_0$ and
consider the implications of the mixing hypothesis on these data. Finally we shall
see that the emerging data from DA$\Phi$NE, presented at this conference,
are in remarkable agreement with these predictions and add weight to
the idea that these scalar mesons are compact $qq\bar{q}\bar{q}$ states with
an extended meson-meson cloud ``molecular" tail. 

\subsection{$f_0$ $a_0$ mixing in central production}

Fig~6c) and d) show the $x_F$ distributions for the 
$a_0^0(980)$ and $a_2^0(1320)$ formed in $p p \to p p a_{0,2}$. The distribution for 
the $a_2^0(1320)$ is similar to that observed for the $a_2^-(1320)$
whereas that for the $a_0^0(980)$ is significantly different and peaks at
$x_F = 0$. 
Indeed this is the only state with $I=1$ that is observed
to have a $x_F$ distribution peaked at zero~\cite{sumpap},
and moreover the distribution for the $a_0^0(980)$ 
looks similar to  the central production of states that are accessible to
\pom \pom \thinspace fusion, 
in particular \pom \pom $\rightarrow f_0(980)$.
%, see
%fig~6c) and d).
\par

\begin{figure}
\epsfxsize200pt
\figurebox{200pt}{160pt}{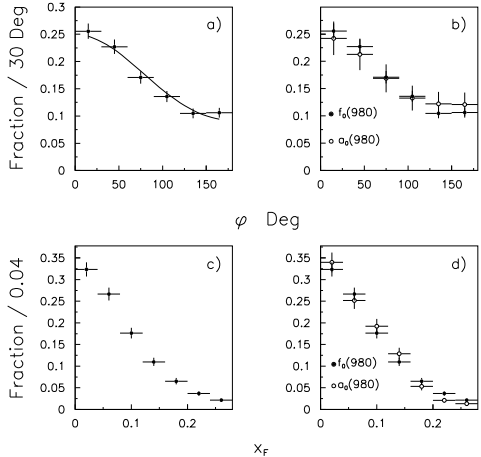}
\vspace{-.5cm}
\caption{The $\phi$ distributions
a) for the reaction $pp \rightarrow pp f_0(980)$ and
b) for the $f_0(980)$ compared to the $a_0^0(980)$. 
The 
$x_F$ distributions
c) for the reaction $pp \rightarrow pp f_0(980)$ and
d) for the $f_0(980)$ compared to the $a_0^0(980)$.}
\end{figure}

In the process  $pp \rightarrow p (\eta \pi^0) p$,
\pom \pom \thinspace fusion 
 will feed only $I=0$ channels, such as the $f_0(980)$ and $f_2(1270)$;
one would not expect this to affect $a_{0,2}$ production 
unless isospin is broken.
The $a_0(980)/a_2(1320)$ ratio in the WA102 data
is relatively large and the $x_F$ distribution of the $a_0(980)$
production is, within the errors, identical to that of the
$f_0(980)$ (see fig.~6d) and
 the $\phi$ distribution for the $a_0(980)$ also looks
very similar to that observed for the $f_0(980)$~(fig.~6b).
Qualitatively this is what would
be expected if 
part of the centrally produced $a_0^0(980)$ is due to
\pom \pom $\to f_0(980)$ followed by
mixing between the $f_0(980)$ and the $a_0(980)$.
\par
Ref \cite{ClK01b} found
 that 80~$\pm$~25~\% of the $a_0^0(980)$ comes from the
$f_0(980)$ and upon
combining this result with the relative total cross sections for the production
the $f_0(980)$ and $a_0^0(980)$~\cite{sumpap} they found the
$f_0(980)-a_0(980)$ mixing intensity to be 8~$\pm$~3~\%.
\par
 This adds weight to the hypothesis that the 
$f_0(980)$ and $a_0(980)$
are siblings that strongly mix, and that the $a_0(980)$ is not simply
a $^3P_0 q\bar{q}$ partner of
the $a_2(1320)$. This is consistent with the $0^{++}(qq\bar{q}\bar{q}/MM)$ picture
of these states and
a natural explanation of these results is that
$K\bar{K}$ threshold plays an essential role in the existence and 
properties.
\par
Other lines of study are now warranted.
Experimentally to confirm these ideas requires measuring the
production of the $\eta\pi$ channel at a much higher energy, for
example, at LHC, Fermilab 
or RHIC where any residual Reggeon exchanges such as $\rho \omega$
would be effectively zero
and hence any $a_0(980)$ production must come from isospin breaking
effects. On the theory side, detailed predictions are needed in specific
models in order to resolve precisely how the $K\bar{K}$ threshold
relates to the $f_0(980)/a_0(980)$ states.
\par
Other ``pure" flavour channels should now be
explored. Examples are $D_s$ decays
%~\cite{lipkin}
where the weak decay leads to a pure I=1 light hadron final state. Thus 
$\pi f_0(980)$ will be 
(and is~\cite{e791}) prominent, while the mixing results suggest
that $\pi a_0$ should also be present at $8 \pm 3$ \% intensity. Studies
 with high statistics data sets now emerging
from E791, Focus and BaBar are called for, and also
 studies of $J/\psi$ decays at Beijing,
 in particular to the
``forbidden" final states $\omega a_0$ and $\phi a_0$ where ref\cite{ClK01b}
 predicts
branching ratios of $O(10^{-5})$.

\subsection{$\phi \to \gamma f_0/\gamma a_0$}

The radiative decays of the $\phi \to \gamma f_0(980)$ and
$\gamma a_0(980)$ have long been recognised as a potential route towards disentangling 
their nature. 
In this section I note that isospin mixing effects could considerably alter
some predictions in the literature for $\Gamma(\phi \to \gamma f_0(980))$ and
$\Gamma(\phi \to  \gamma a_0(980))$, and I show that new data from DA$\Phi$NE
promise to reveal their nature.

The magnitudes of these widths are predicted to be rather 
sensitive to the fundamental structures
of the $f_0$ and $a_0$, and as such potentially discriminate amongst them.
For example, if $f_0(980) \equiv s\bar{s}$ and the dominant dynamics is
the ``direct" quark transition
$\phi(s\bar{s}) \to \gamma 0^{++}(s\bar{s})$, then
 the predicted $b.r.(\phi \to \gamma f_0)
\sim 10^{-5}$, the rate to $\phi \to \gamma a_0(q\bar{q})$ being even 
smaller due to OZI supression\cite{cik}.
 For $K\bar{K}$ molecules the rate was predicted\cite{cik}
to be higher, $\sim (0.4 - 1) \times 10^{-4}$, while for tightly compact
$qq\bar{q}\bar{q}$ states the rate is yet higher\cite{achasov3,cik}, $\sim 2 \times 10^{-4}$.
Thus at first sight there seems to be a clear means to distinguish amongst them.

In the $K\bar{K}$ molecule and $qq\bar{q}\bar{q}$ scenarios 
it has uniformly been assumed that the radiative transition will be
driven by an intermediate  $K^+K^-$ loop ($\phi \to K^+K^-
\to \gamma K^+K^-  \to \gamma 0^{++}$). Explicit calculations
in the literature agree that this implies\cite{achasov3,cik,cbrown,lucieu}

\begin{equation}
b.r. (\phi \to f_0(980)\gamma ) \sim 2 \pm 0.5 (10^{-4}) \times F^2(R)
\end{equation}

\noindent where $F^2(R) = 1$ in point-like effective field theory
computations, such as refs.\cite{achasov3,lucieu}. (The range of
predicted magnitudes for the branching ratios reflects 
the sensitivity to assumed parameters, such as masses and couplings
that vary slightly among these references). 
By contrast, if the $f_0(980)$ and $ a_0(980)$ are spatially extended $K\bar{K}$
molecules, (with r.m.s. radius $R > O(\Lambda_{QCD}^{-1}$)),
then  the high momentum region of the integration 
in  refs.\cite{cik,cbrown} is cut off, leading in effect
to a form factor suppression, $F^2(R) < 1$\cite{cik,achasov4}.
The differences in absolute rates are thus intimately linked to the
model dependent magnitude of $F^2(R)$.

Precision data
on both $f_0$ and $a_0$ production are now available  from DA$\Phi$NE\cite{kloe}.
Before discussing this, there are two particular items that I wish
to address concerning the current predictions.
One concerns the absolute branching ratios, and the second concerns the
ratio of branching ratios where,
if $f_0$ and $a_0^0$ have common constituents (and hence are ``siblings")
and are eigenstates of isospin, then their
affinity for $K^+K^-$ should be the same and so\cite{achasov3,cik,lucieu}

\begin{equation}
\frac{\Gamma(\phi \to  f_0 \gamma)}{\Gamma(\phi \to  a_0 \gamma)} \sim 1.
\end{equation}

 There are reasons to be suspicious of the predictions in both eqs. (1) and (2).
I frame these remarks in the context of the $K\bar{K}$ molecule,
but they apply more generally.

If in the $K\bar{K}$ molecule one has

\begin{equation}
|f_0\rangle = cos\theta |K^+K^-\rangle + sin\theta |K^0\bar{K^0} \rangle 
\end{equation}
\noindent and
\begin{equation}
|a_0^0\rangle = sin\theta |K^+K^-\rangle - cos\theta |K^0\bar{K^0} \rangle
\end{equation}

\noindent then the branching ratios $\phi \to \gamma f_0(\gamma a_0)$
 as found in ref.\cite{cik}
can be summarised as follows

\[
b.r.(\phi \to \gamma f_0:\gamma a_0) =  
(4 \pm 1)
\times 10^{-4}(cos^2\theta:sin^2\theta)
\]
\[ \times 
(\frac{g^2_{SK^+K^-}/4\pi}{0.58GeV^2}) F^2(R) 
\]

\noindent As shown in ref.\cite{cik}, the analytical results of point-like effective field
theory calculations (e.g. refs.\cite{achasov3,lucieu}) can be recovered
as $R \to 0$, for which
$F^2(R) \to 1$. In contrast to the compact hadronic
four quark state, the $K\bar{K}$ molecule is
spatially extended with r.m.s. $R \sim 1/\sqrt{m_K \epsilon}$, where $\epsilon$
is the binding energy and $F^2(R) < 1$,
 the precise magnitude depending on the $K\bar{K}$
molecular dynamics.

It is clear also that the predictions of the absolute rates above 
are driven by (i) the assumed
value for $\frac{g_{SK^+K^-}^2}{4\pi}  = 0.58$ GeV$^2$, and
(ii) the further assumption that the $f_0$ and $a_0$ are
$K\bar{K}$ states with $I=0,1$: hence 
 $\theta = \pi/4$.

There is some uncertainty about the former, which needs to be
experimentally studied further. 
In particular, an analysis of Fermilab E791\cite{e791}  data,
which studies the $f_0(980)$ produced in $D_s$ decays, even suggests that
$\frac{g_{fK^+K^-}^2}{4\pi}  \sim 0.02 \pm 0.04 \pm 0.03$ (GeV)$^2$, hence
consistent with zero! However,it should be noted that only the
$\pi \pi$ decay mode of the $f_0(980)$ has been studied in this experiment
and hence the coupling to $K^+K^-$ is only measured indirectly.
 With such uncertainties in the
value of this coupling strength, predictions of absolute rates for 
$\phi \to \gamma f_0(980)$ or $\phi \to \gamma a_0(980)$ via an intermediate
$K\bar{K}$ loop must be treated with some caution.
By contrast, in the ratio of branching ratios this uncertainty is reduced,
at least in the case of $K\bar{K}$ molecules for which\cite{cik}
$\frac{\Gamma(\phi \to \gamma f_0)} {\Gamma(\phi \to \gamma a_0)} \sim 1$.

The central production, discussed above, suggested that there is a significant
mixing. Specifically: 
  in 
(isoscalar) \pom (Pomeron)-induced 
production in the central region at high energy,
production of the $a_0^0(980)$ comes dominantly from mixing with the
$f_0(980)$ such that the $f_0 - a_0$ are not good isospin eigenstates.
In the language of the $K\bar{K}$ molecule, at least, this
would translate into  $\theta \neq \frac{\pi}{4}$ 
and hence to a
 significant difference in behaviour for $\Gamma(\phi \to \gamma f_0)/
\Gamma(\phi \to \gamma a_0)$.

With the basis as defined in eqs.~(3) and (4), 
the ratio of production rates by \pom \pom (isoscalar)
fusion in central production will be
$
\sigma (\pome \pome \to a_0)/\sigma (\pome \pome \to f_0)
 = \frac{1-sin 2\theta}{1+sin 2 \theta}
$

\noindent Ref.\cite{ClK01b} found this to be $ (8 \pm 3) \times 10^{-2}$.
Hence if we assume that the production phase is the same for the two,
then  within this approximation
the relative rates are predicted to be\cite{ClK01c}
\begin{equation}
\frac{\Gamma(\phi \to \gamma f_0)}{\Gamma(\phi \to \gamma a_0)} \equiv cot^2\theta 
= 3.2 \pm 0.8
\end{equation}

\noindent This is far from the naive expectation of unity for ideal isospin states and is 
 a rather direct consequence of the isospin mixing obtained
in ref.~\cite{ClK01b}. In order to use the data to abstract magnitudes
of $F^2(R)$, and hence assess how compact the four-quark
state is,a definitive accurate value for $g_{fKK}^2/4\pi $ will be required.
If for orientation we adhere to the value used elsewhere, $g_{fKK}^2/4\pi\sim 0.6$
GeV$^2$,
 and impose the preferred $\theta$,
then the results of ref.~\cite{cik} are revised to

\begin{equation}
b.r.(\phi \to \gamma f_0) + b.r.(\phi \to \gamma a_0) \leq (4 \pm 1) (10^{-4}) 
\end{equation}

\noindent and 

\begin{equation}
b.r.(\phi \to \gamma f_0) =  (3.0 \pm 0.6)  10^{-4} F^2(R)
\end{equation}
\begin{equation}
b.r.(\phi \to \gamma a_0) =  (1.0 \pm 0.25) 10^{-4} F^2(R)
\end{equation}

Refs\cite{barnes} and \cite{cik} 
developed a simple potential picture of a
$K\bar{K}$ molecule which led to 
$R \sim 1.2 fm$, $F^2(R) \sim 0.25$.
However, the predictions are rather sensitive to the assumed details and
more sophisticated treatments, including 
mixing between $K\bar{K}$ and $qq\bar{q}\bar{q}$ are now warranted.

It is therefore most interesting that data presented here
find \cite{kloe}

\begin{equation}
\frac{\Gamma(\phi \to \gamma f_0)}{\Gamma(\phi \to \gamma a_0)} \equiv cot^2\theta 
= 4.1 \pm 0.4
\end{equation}

\noindent consistent with  the predicted ratio in eq.~(6). The 
individual rates may therefore be used as a measure of $F^2(R)$. Branching ratios for 
which $F^2(R) << 1$ would imply that the $K^+K^-0^{++}$
interaction is spatially extended, $R > O(\Lambda_{QCD}^{-1})$. Conversely,
for $F^2(R) \to 1$, the system would be spatially compact, as in $qq\bar{q}\bar{q}$. 

 The data from KLOE are \cite{kloe}

\begin{equation}
b.r.(\phi \to \gamma f_0) =  (2.4 \pm 0.1)  10^{-4} 
\end{equation}
\begin{equation}
b.r.(\phi \to \gamma a_0) =  (0.6 \pm 0.05) 10^{-4} 
\end{equation}

\noindent which, compared with eqs.(7,8), imply $F^2(R) \sim 0.7 \pm 0.2$, supporting
the qualitative picture  of a compact $qq\bar{q}\bar{q}$ structure 
that spends a sizeable part of its lifetime in a two meson state, such
as $K\bar{K}$. An important feature of this also is that there is a significant
isospin mixing at work, driven by the $K^+K^-$-$K^0\bar{K^0}$ mass difference.
A subtle and unexplained
violation of isospin has already been noted\cite{bramon} in the ratio
$\phi \to K^+K^-/K^0\bar{K^0}$, whose origin may 
also touch on these issues.

\section {The $\xi(2.2)$}

There is a tantalizing signal that has been claimed\cite{beszeta}
as evidence for a tensor glueball. I have severe doubts about this state, but first let me
present the ``case for the prosecution". It is narrow ($\sim 20$MeV)
and seen in a glueball favoured production channel: $\psi \to \gamma \xi$
where $\xi \to K^+K^-;K^oK^0,\pi^+\pi^-,p\bar{p}$, according to BES\cite{beszeta}.
In each channel the effect is $\sim 4\sigma
$ 
but the actual number of events is small.
Support is claimed from old data by Mark3 at SPEAR\cite{mk3zeta}
where a similar structure was seen in $K\bar{K}$.
 However, closer examination begins to reveal questions
that merit further study. 

First, DM2 \cite{DM2} see no evidence for a narrow state in either $K^+K^-$ or $K_S^0K_S^0$.
Note also that Mark3 data\cite{mk3zeta} on the $p\bar{p}$ (marginally) and $\pi\pi$ (more significantly) 
do not add support for this state. But let's press on in hope. Two
further pieces of data have been used to support the claim that $\xi$
is a glueball. First, from LEAR one has no signal in $p\bar{p} \to \xi \to K\bar{K}$
%leading to an upper limit of 1.5$\times 10^{-4}$ for this branching ratio
\cite{LEARzeta},
which when combined with the signal for each of the individual
$\xi \to K\bar{K}; \xi \to p\bar{p}$ from BES implies a large intrinsic
production: $br.(\psi \to \gamma \xi) > 10^{-3}$. Such a magnitude
would be in line with expectations for a tensor glueball if $\Gamma_{tot}
\sim 20$MeV\cite{cfl}, but at the price of having detected only a limited
fraction of the decay channels.
A further piece of evidence has been presented to this conference from L3.
They find no signal in $\gamma \gamma$ and place an upper limit:
$\Gamma(\xi \to \gamma \gamma)b.r.(\xi \to K_sK_s) < 1.4eV$\cite{L3}.

So, we appear to have a large production in the glueball friendly
$\psi$ decay and a strong suppression in the ``anti-glueball" $\gamma \gamma$
channel, which leads some to assert that
the $2^{++}$ glueball is the $\xi$. The problem, in my opinion, is that the claims for glueball are based
on what is $\bf{not}$ being seen! Furthermore, there is no convincing signal for $\xi$ in
any other experiment. There is another possible interpretation of the 
LEAR and L3 non-observations: the $\xi$ does not exist!

If it does exist, another solution is that the $K\bar{K}$ decay is dominant (as in
the Mark 3 data) and that, in aprticualr, $p\bar{p}$ is suppressed. 
The LEAR limit would then be less restrictive, and allow 
$br.(\psi \to \gamma \xi) \sim 10^{-4}$. The true $\xi$ could then be a broader state
( as seen elsewhere, for example WA102\cite{WAkk}, DM2\cite{DM2} and \cite{GKI84}) and with 
$br.(\psi \to \gamma \xi) \sim 10^{-4}$ be compatible with a $q\bar{q}$
state\cite{cfl}. If this state has a significant $s\bar{s}$ content, an excited
tensor with width $\sim 100 MeV$, then it is probably
compatible with the L3 $\gamma \gamma$ limit.

This is clearly a question that needs to be pursued with high priority
at an upgraded BES and at a downgraded (in an energy sense!) CESR. High statistics
and independent analyses should determine whether this state is real or not. 
If it is confirmed with the above properties then it will be rather compelling;
however, at the present time I am of the opinion that
absence of evidence may be evidence of absence. The challenge for
the new facilities will be to settle this question.

\section{Summary and Prospects} 
 
Establishing that gluonic degrees of freedom are being excited is now
a real possibility and centres on understanding (a) the scalar mesons (b)
being able to distinguish between hybrids and molecules
or compact four-quark states for exotic $J^{PC}=1^{-+}$.

A``pure" molecule would be produced and decay in the mesons that make
it; a compact four quark or hybrid would show up in channels driven by the flavour
spin clebsches. Furthermore, the latter configurations will be 
in nonets (any other representation would immediately eliminate hybrid)
and the mass dependent pattern will in general be different for
$q\bar{q}$/hybrid and $(qq)^{\bar{3}}(\bar{q}\bar{q})^3$. So, unless
Nature is malicious, it will be possible to answer this puzzle definitively.

 The scalars below 1 Gev are too light to enable a simple
distinction between loose molecules and compact four-quarks states to
be felt in decays (except perhaps for the $a_0(980)$ where the $K\bar{K}$
and $\eta\pi$ both couple strongly and point to a significant compact
four-quark feature). However, the production dynamics and systematics of
these states is interesting and full of enigmas, which may be soluble if
one adopts the four-quark/molecule picture.

$D_s$ decays into $\pi f_0$ clearly point to an $s\bar{s}$ presence in
the $f_0$. However, the production in Z decays is rather
non-strange-like\cite{lafferty}.
$\psi$ to $\omega f_0$ and $\phi f_0$ also do not equate easily with a
simple $q\bar{q}$ description for $f_0$ and $a_0$. 
 The central production in pp shows that $f_0$ is strongly
produced, akin to other $n\bar{n}$ states and much stronger than
$s\bar{s}$ which appear to be suppressed in this mechanism. Furthermore,
$f_0$
survives the $dk_T \to 0$ filter of ref.\cite{ck97}.
 The systematics of this
filter I believe (and would like to prove) is driven by S-wave
production: this would be fine for either a compact four-quark or
molecule. None of these phenomena fit easily with an intrinsic $3P_0$
$q\bar{q}$ as the dominant constituency.

Whenever S-wave dynamics can play a role it will overide P-waves; so
one expects $K\bar{K}$ S-wave production to drive the $f_0/a_0$ whenever
allowed. This is indeed what happens in the $\phi \to \gamma f_0/\gamma a_0$;
the``large" rate cries out for the $K^+K^-$ loop to drive it. A question
is whether the $s\bar{s} n\bar{n}$ constituents of the intermediate state``between" the 
initial $\phi$ and final $f_0$ are
able to fluctuate spatially enough to be identified as two colour
singlet K's, which then couple to the $f_0$, or whether they are a
compact system in the sense of being confined within $\sim$ 1fm. The
former would have some form factor suppression of the rate; the latter
would be more pointlike and larger rate. The emerging data are
between these extremes, but nearer to the expectations for a compact configuration. 

Knowledge on the $\gamma \gamma$ couplings is lacking and better 
data would be useful. We
know that for the $2^{++}$ $\gamma \gamma$ reads the compact $q\bar{q}$
flavours; there is no 2-body S-wave competition in the imaginary part as
$\rho\rho$ etc are too heavy. I would expect that for the $0^{++}$ the
$K\bar{K}$ will dominate the $\gamma \gamma$ if there is a KK component
in the wavefunction. At the other extreme; were the state a pure compact
four quark, then higher intermediate states - KK, KK*,KK** etc - would
all be present. Achasov\cite{achasov3} has discussed these
and a precise calculation has many problems, but the
{\bf ratios} of $\gamma \gamma$ to $f_0/a_0$ would probably be sensitive and
more reliable. 

The theoretical frontier suggests that
 one can divide the phenomenolgy of scalars into those above and
those below 1 GeV. I suspect
that much of the confusion will begin to evaporate if one adopts such a
strating point. Empirically,signs of gluonic excitation are
appearing (i) in the form of hybrids with the exotic $J^{PC} = 1^{-+}$
now seen in various channels and more than one experiment (ii) with
$0^{-+}$ and $1^{--}$ signals in the 1.4 - 1.9GeV region that do not
fit well with conventional quarkonia and show features predicted for hybrids
(iii) in the form of the scalar glueball mixed in with quarkonia in the 1.3 - 1.7 Gev 
mass range. Theoretical questions about the latter are concerned with whether its effects
are localised above 1GeV, or whether they are spread across a wider mass range, even
down to threshold. Experimental questions that need to be resolved concern the
existence and properties of the $f_0(1370)$ and $a_0(1400)$.

These questions in turn provoke my list of challenges for experiment.

(i) In $e^+e^-$, the region of hybrid charmonium promises to be $\sim 4GeV$.
This would be an excellent area for study at an upgraded BES and the
proposed CESR Tau-Charm-Factories. At lower energies, VEPP and DA$\Phi$NE
can study the 1.4-1.7GeV region where light flavour hybrid vectors
may occur. High statistics studies of
radiative decays of such states into the $f_0(980);f_1(1285);
f_2(1270)$ could teach us much\cite{ClDK01}. HERA can also investigate the
production mechanism, $Q^2$ dependence etc of the vectors in hope
of distinguishing radial excited quarkonia from hybrids.

(ii) Photo and electroproduction of hybrids $\sim 2$GeV mass can be
studied at Jefferson Lab, via $\pi$ and $\omega$ exchange. The
vector nature of the photon, and its mixed isospin content, can access
the exotic quantum numbered ``golden" hybrids, such as $1^{-+}$. In any
event, the properties of the newly-sighted $1^{-+}$ around 1.4 - 1.6Gev
should be investigated. The
moderate energies available here can actually be an aid. 

(iii) $\gamma \gamma$ couplings give rather direct information on the
flavour content of C=+ states. Such information on the scalar mesons
will be an essential part of interpreting these states.

(iv) Heavy flavour decays, in particular $D_s$ and $D$ into $\pi$ and associated
hadrons can access the scalar states. Precision data are needed to disentangle the
contributinos of the various diagrams, whereby the flavour content of the scalars
can be inferred. There is also a tantalising degeneracy between the $\pi_g(1.8)$ and
the $D$, which may radically affect the Cabibbo suppressed decays of the latter.
Hence precision data on such charm decays is warranted.

(v) Central production need now to be studied at higher energies, at RHIC,
Fermilab and possibly the LHC. The $0^{-+}$ glueball should be manifested once the
$t_1t_2 \to 0$ kinematic suppression is overcome. The systematics of
scalar and tensor production, with their $\phi$ and $dk_T$ dependences, need
to be understood.

(vi) Tau Charm Factories may at last appear. Obtaining a well defined universally 
accepted data-set on $\psi$ decays is needed; a problem for phenomenology is that
data from different experiments do not always agree. Clarifying the status
of $\xi(2.2)$ is needed from independent experiments, such as BES and CESR should
provide. Finally, $\chi$ decays offer an entree into light flavoured states; the
excitement about the scalar glueball mixing with the quarkonia nonet began when the
precision data from $p\bar{p}$ annihilation at LEAR first emerged. Data at rest
were beautiful and well analysed. Data in flight however tend to be more problematic,
not least as one cannot so easily control knowledge of the incident partial wave.
$\chi$ decays can access these phenomena, at c.m. energies up to 3.5GeV, and from well
defined initial $J^{PC}$ states. In particular, the $1^{++}$ decay into $\pi + X$
probes $X \equiv 1^{-+}$ in S-wave.  

I hope that this talk has shown some ways in which current and future experiments
can impact on the physics of light flavours, and of glue. A TCF promises
to be an important feature in this field; hopefully, together
with BEPCII, That's Cornell's Future.

\begin{center}
{\bf Acknowledgements}
\end{center}
\par
This is supported, in part, by 
the EU Fourth Framework Programme contract Eurodafne, FMRX-CT98-0169.

\end{document}